\documentclass[twocolumn]{aastex62}
\usepackage{amsmath,amssymb,graphicx,nccmath}
\usepackage{natbib}
\setcitestyle{notesep={; }}
\received{December 13, 2017}
\accepted{July 19, 2018}

\submitjournal{ApJ}

\shorttitle{MN 90 and HD 168625}
\shortauthors{Arneson et al.}

\begin{document}
\title{SOFIA/FORCAST Observations of the Luminous Blue Variable Candidates MN 90 and HD 168625}

\correspondingauthor{Ryan Arneson}
\email{ContactRyanArneson@gmail.com}

\author{Ryan A. Arneson}
\affil{Minnesota Institute for Astrophysics, School of Physics and Astronomy, University of Minnesota, 116 Church Street S.E., Minneapolis, MN 55455, USA}

\author{Dinesh Shenoy}
\affil{Minnesota Institute for Astrophysics, School of Physics and Astronomy, University of Minnesota, 116 Church Street S.E., Minneapolis, MN 55455, USA}

\author{Nathan Smith}
\affil{Steward Observatory, University of Arizona, 933 North Cherry Avenue, Rm 336, Tucson, AZ 85721, USA}

\author{Robert D. Gehrz}
\affil{Minnesota Institute for Astrophysics, School of Physics and Astronomy, University of Minnesota, 116 Church Street S.E., Minneapolis, MN 55455, USA}

\newcommand{\massrate}{$M_\sun \, yr^{-1}\:$}
\newcommand{\kms}{$\rm{km \, s^{-1}}~$}
\newcommand{\solarmass}{$M_\sun\:$}
\newcommand{\twodust}{{\bf{2-D}}ust}
\newcommand{\AKs}{$A_{K_s}$}

\begin{abstract}
We present SOFIA/FORCAST imaging of the circumstellar dust shells surrounding the luminous blue variable (LBV) candidates MN 90 and HD 168625 to quantify the mineral abundances of the dust and to constrain the evolutionary state of these objects.  Our image at 37.1 \micron\ of MN 90 shows a limb-brightened, spherical dust shell.  A least-squares fit to the spectral energy distribution of MN 90 yields a dust temperature of $59 \pm 10$ K, with the peak of the emission at 42.7 \micron.  Using \twodust\ radiative transfer code, we estimate for MN 90 that mass-loss occurred at a rate of $(7.3\pm0.4) \times10^{-7}$ \massrate $\times\ (v_{exp}/50\, $\kms\!) to create a dust shell with a dust mass of $(3.2\pm0.1)\times10^{-2}$ \solarmass.  Our images between 7.7 -- 37.1 \micron\ of HD 168625 complement previously obtained mid-IR imaging of its bipolar nebulae.  The SOFIA/FORCAST imaging of HD 168625 shows evidence for the limb-brightened peaks of an equatorial torus.  We estimate a dust temperature of $170 \pm 40$ K for the equatorial dust surrounding HD 168625, with the peak of the emission at 18.3 \micron.  Our \twodust\ model for HD 168625 estimates that mass-loss occurred at a rate of $(3.2\pm0.2)\times10^{-7}$ \massrate to create a dust torus/shell with a dust mass of $(2.5\pm0.1)\times10^{-3}$ \solarmass.
\end{abstract}


\section{Introduction}
\label{intro}
It is widely recognized that luminous blue variables (LBVs) represent a post-main sequence phase in which massive stars (initial mass $M_i \geq$ 20 \solarmass; \citealp{Langer94}) lose a considerable amount of mass via giant eruptions and minor outbursts.  From the expansion velocities of known LBV nebulae, a dynamical age of a few $10^4$ years is usually inferred, which points to a very short-lived evolutionary phase -- only about 40 are known \citep{Clark05, Weis11, Naze12}.  Although the category is still not unambiguously defined, these objects generally exhibit a high luminosity ($\geq 10^{5.5}$ $L_\sun$), low amplitude photometric variability ($\sim$ 0.1 mag) on timescales ranging from weeks to months, and a larger, irregular photometric variability, called S Dor variability, with amplitudes of 1 -- 2 mag occuring on timescales of years to decades with mass-loss rates of $\sim 10^{-5} - 10^{-4}$ \massrate\!.  In addition, some LBVs exhibit giant eruptions, $\eta$ Car  being the most famous example \citep{Humphreys99}.  These giant eruptions are responsible for producing circumstellar nebulae with sizes up to 1 -- 2 pc and expansion velocities anywhere from 10 \kms to several hundred \kms for most LBVs \citep{Weis11} that are then shaped by wind-wind interactions \citep{vanMarle07}.  In rare cases, very fast speeds have been seen, reaching as high as 6000 \kms in the case of $\eta$ Car \citep{Smith08}.  These nebulae can have various morphologies; \citet{Weis11} estimates 50\% are bipolar, 40\% spherical, and 10\% are irregular.  The bipolar nebulae may be formed by density gradients in the wind \citep{Frank95}, different mass-loss episodes in which the wind changes from equatorial to polar during the bistability jump \citep{Smith04}, or the rotation of the star \citep{Dwarkadas02, SmithTownsend07}.  Thus, the morphology of the nebula discloses the mass-loss history and evolutionary stage of the central star and the circumstellar environment.

In this work we present 5 -- 40 \micron\ mid-IR observations with the Faint Object infraRed CAmera for the SOFIA Telescope (FORCAST; \citealp{Herter12}) instrument on board the NASA Stratospheric Observatory for Infrared Astronomy (SOFIA; \citealp{Becklin07, Gehrz09, Young12}) of the two compact nebulae, MN 90 and HD 168625.  By imaging these nebulae at a range of wavelengths, we can study their structure, estimate the dust composition, and quantify the dust temperature and total mass. These parameters will help to determine the physical properties of the circumstellar environments and to constrain the importance of eruptive mass-loss in post-main sequence stellar evolution.  In Section \ref{obs} we summarize the observations and data reduction strategies.  Section \ref{model} describes the axisymmetric radiative transfer code {\bf{2-D}}ust and the derived dust geometry parameters and inferred mass-loss histories.  We discuss the results of our analysis in Section \ref{discussion} and present the conclusions in Section \ref{conclusion}.

MN 90 (central star 2MASS J18455593-0308297) was discovered and catalogued by \citet{Gvaramadze10} using the Multiband Imaging Photometer (MIPS; \citealp{Rieke04}) aboard the \emph{Spitzer Space Telescope} \citep{Werner04, Gehrz07} in the MIPS Galactic Plane Survey (MIPSGAL; \citealp{Carey09}), which mapped 278 deg$^2$ of the inner Galactic plane: $-65\degr < l < -10\degr$ and $10\degr < l < 65\degr$ for $|b| < 1\degr$.  It is one of a large number of similar shells found in MIPSGAL images that resemble the circumstellar nebulae of LBVs and late WN-type Wolf Rayet stars (WNL).  The lack of optical counterparts for most of them indicates they are highly obscured \citep{Wachter10}.  Follow-up spectroscopy of some of the other MIPS Nebulae has revealed them to be LBVs, candidate LBVs or early-type supergiants (see the summary list in \citealp{Kniazev15}).  MN 90 appears nearly circular as projected on the sky and appears to be a limb-brightened shell with a star at the center.  The star is undetected in \emph{Swift} UV images, Palomar Observatory Sky Survey (POSS) visual images, and the ground-based visual images used to make the HST Guide Star Catalog.  The distance to MN 90 is unknown and there is little information about its physical parameters.   \citet{Wachter10} included MN 90 along with a large sample of numerous similar shells for follow-up spectroscopy but ultimately did not report any spectra for this particular star.  Based on the analysis of other stars in their sample, they grouped MN 90 along with other shells for which they predict the stars are early-type.  \citet{Naze12} did not dectect any X-ray emission from MN 90 in their XMM-\emph{Newton} survey of a sample of known and candidate LBVs.  \citet{Mizuno10} list MN 90 as MGE029.5086-00.2090, but other than reporting a MIPSGAL flux at 24 \micron, there is nothing more specific about the central star.  We are unaware of any other imaging or spectroscopy of MN 90 that may have been obtained.   Hereafter we will use ``MN 90'' to refer to both the nebula and the central star.

HD 168625 (IRAS 18184--1623) was first identified as a candidate LBV by \citet{Chentsov89}, who classified it as spectral type B5.6 $\pm$ 0.3 with $T_{*} \simeq 13000$ K.  Its spectral type seems to vary from B2 \citep{Popper40} to B8 \citep{Morgan55}, although no dramatic light-variations have been been reported in the last 40 years \citep{vanGenderen92, Sterken99}.  The lack of evidence for large variations kept HD 168625 from being classified as an LBV.  However, it was found to be LBV-like by \citet{Hutsemekers94} who, using near-infrared (near-IR) and visible imaging and spectroscopy, found a high mass-loss rate and a shell with two regions: an inner $10\arcsec \times 13\arcsec$ elliptical ring and a perpendicular outer horn-shaped region suggesting a bipolar outflow.  \citet{Nota96} used deeper H$\alpha$ imaging of the nebula to identify faint filaments in the bipolar structure that extended to $16\arcsec \times 21\arcsec$, indicating that a LBV-like major outburst occurred $\sim 10^3$ years ago.  Given its curious characteristics, HD 168625 has been the focus of several additional studies.  \citet{Meixner99} included it in their large, mid-infrared (mid-IR) proto-planetary nebula candidates survey and imaged it at 8.8, 12.5, and 20.6 \micron\ revealing a toroidal dust shell.  From mid-IR images at 4.7, 10.1, 11.6, and 19.9 \micron, \citet{Robberto98b} used an analytical spherical model to derive a dust temperature of 135 K and a dust mass of 2.8$\times10^{-3}$ \solarmass.  Similarly, \citet{Pasquali02} used mid-IR imaging at 4 and 11 \micron\ to derive a dust temperature of 113 K, and a nebular expansion velocity of 19 km s$^{-1}$ from ground-based echelle spectra.  \citet{OHara03} modeled the morphology of the toroidal dust region using {\bf{2-D}}ust and found a dust mass of $2.5 \pm 0.1 \times 10^{-3}$ \solarmass$\!$.  \citet{Mahy16} used far-infrared (far-IR) imaging and spectroscopy with optical spectra to constrain the CNO abundances of HD 168625 and the surrounding nebula to determine that the central star had an initial mass between 28 -- 33 \solarmass\ and lost its material after the blue supergiant phase.  \citet{Smith07}
presented \emph{Spitzer} images showing that the bipolar lobes and torus of HD 168625 were actually a triple-ring system that closely resembled the ringed nebula around SN 1987A. He pointed out that a single rotating star could potentially eject an equatorial torus even if it is not rotating at the critical speed.  Presumably this can occur because the stars approach or violate the classical Eddington limit during their giant eruptions when the mass is ejected (see, e.g., \citealp{SmithTownsend07}) allowing the star's rotation to be more influential at the resulting lower effective gravity.  \citet{Taylor14} found no evidence for a binary companion to HD 168625 from radial velocity monitoring and a modest rotation speed of 53 \kms\!.  The presence of a possible companion star was detected by \citet{Aldoretta15} using interferometric observations.  A wide-orbit companion with a projected separation of 1.15\arcsec\ was later confirmed by \citet{Martayan16} using adaptive optics images, but there is no evidence that it is gravitationally bound to HD 168625 and it is too distant to have any impact on the shaping of the nebula.  The role companions might play in the formation and evolution of LBV and LBV-like objects is still not well understood. 

\section{Observations \& Data Reduction}
\label{obs}
The targets were observed with SOFIA/FORCAST during Guest Investigator (GI) Cycles 2 and 3.  Descriptions of the SOFIA Observatory and its science instrument (SI) suite have been given by \citet{Becklin07}, \citet{Gehrz09}, and \citet{Young12}.

FORCAST is a dual-channel mid-IR imager covering the 5 -- 40 \micron\ range.   Each channel uses a 256 $\times$ 256 pixel array and provides a distortion-corrected 3.2$\arcmin$ $\times$ 3.2$\arcmin$ field of view with a scale of 0.768$\arcsec$ pix$^{-1}$.   The Short Wave Camera (SWC) uses a Si:As blocked-impurity band (BIB) array optimized for $\lambda < 25$ \micron, while the Long Wave Camera's (LWC) Si:Sb BIB array is optmized for $\lambda > 25$ \micron.  Observations were taken in standard two-position chop-and-nod mode with the direction of the nod matching the direction of the chop (NMC).  The data were reduced by the SOFIA Science Center using the FORCAST Redux v1.0.1$\beta$ and v1.0.6 pipelines \citep{Clarke15} for HD 168625 and MN 90, respectively.  After correction for bad pixels and droop effects, the pipeline removed sky and telescope background emission by first subtracting chopped image pairs and then subtracting nodded image pairs.  The resulting positive images were aligned and merged.  The merged images were then coadded using a robust mean.

We observed MN 90 on UT 2015 July 3 during Guest Investigator (GI) Cycle 3 using the F371 filter ($\lambda_{0}$ = 37.1 \micron, $\Delta\lambda$ = 3.3 \micron).  The total coadded exposure time for the observation of MN 90 was 1553 sec (25.9 min).  Table \ref{tab:phot} summarizes the observed flux from MN 90.  Observations of the asteroid Vesta provided the flux calibration and PSF, with a near-diffraction-limited FWHM at 37.1 \micron\ of 3.6$\arcsec$.

We observed HD 168625 on UT 2014 June 13 during Guest Investigator (GI) Cycle 2 using eight different filters which are summarized in Table \ref{tab:HDobs} and \ref{tab:phot}.  Observations of Beta Andromedae provided the flux calibration and PSF.

\begin{deluxetable}{ccccc}
\tablecaption{Summary of SOFIA/FORCAST Observations of HD 168625 \label{tab:HDobs}}
\tablewidth{.5\textwidth}
\tabletypesize{\footnotesize}
\tablecolumns{5}
\tablehead{\colhead{Filter} & \colhead{$\lambda_0$} & \colhead{$\Delta\lambda$} & \colhead{Exp. Time} & \colhead{PSF FWHM}\\
\colhead{} & \colhead{(\micron)} & \colhead{(\micron)} & \colhead{(s)} & \colhead{(arcsec)}} 
\startdata
F077 & 7.7 & 0.47 & 30 & 2.5\arcsec \\
F111 & 11.1 & 0.95 & 30 & 2.5\arcsec \\
F197 & 19.7 & 5.5 & 16 & 2.7\arcsec \\
F253 & 25.3 & 1.86 & 25 & 3.0\arcsec \\
F315 & 31.5 & 5.7 & 25 & 3.1\arcsec \\
F336 & 33.6 & 1.9 & 31 & 3.1\arcsec \\
F348 & 34.8 & 3.8 & 30 & 3.4\arcsec \\
F371 & 37.1 & 3.3 & 26 & 3.5\arcsec \\
\enddata
\end{deluxetable}

\subsection{Spectral Energy Distributions}
\label{sed}
\begin{figure*}
\plotone{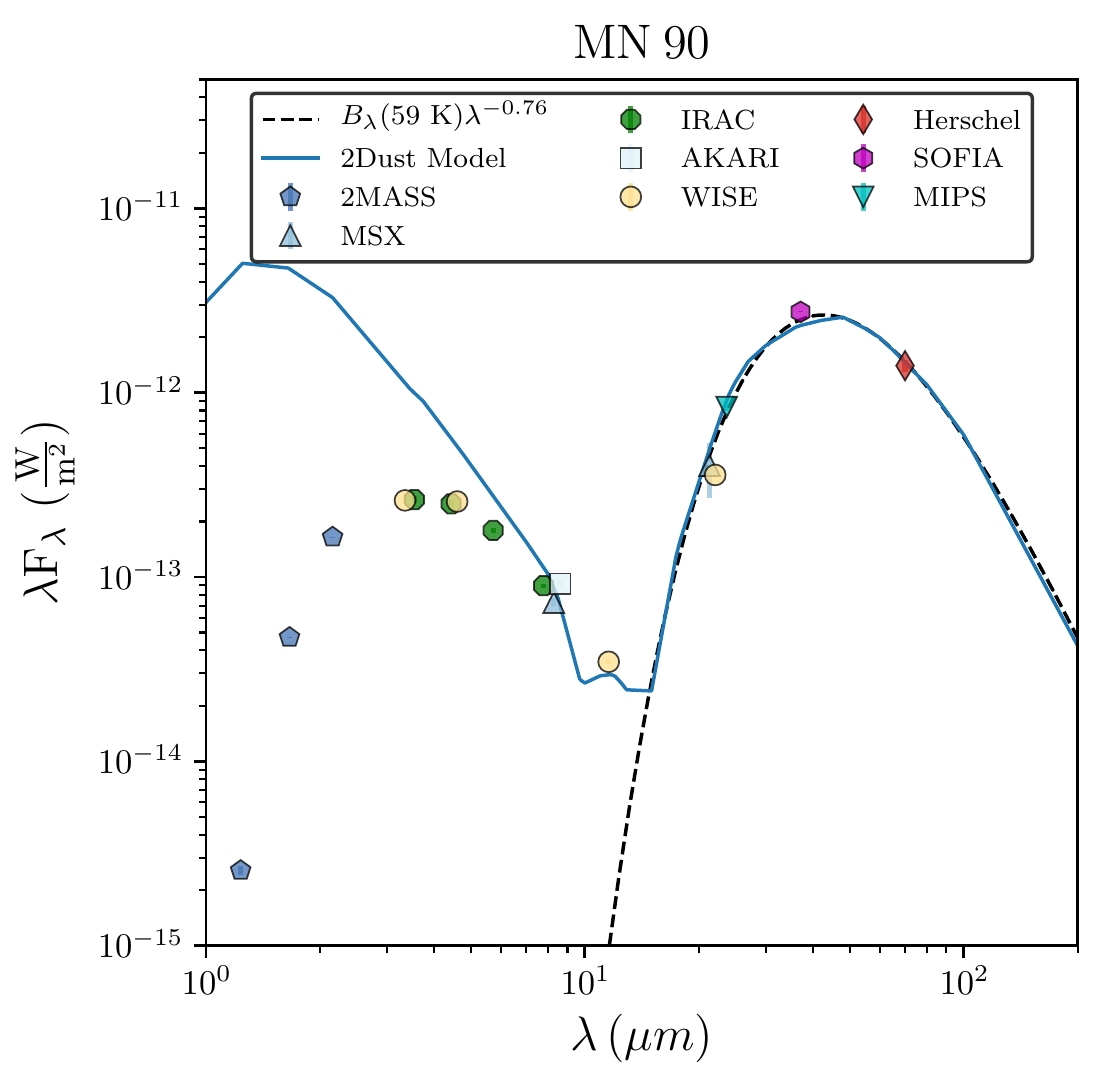}
\figcaption{\label{MN_sed}Observed and model SEDs of MN 90.  The reddened model SED is shown as a solid blue line.  The dashed lines are the best fitting $Q_{\lambda}\cdot B_{\lambda}(T_{d})$ functions with an assumed power law emissivity $Q_{\lambda} \propto \lambda^{-\beta}$, which yields $T_{d}$ = $59 \pm 10$ K, $\beta$ = $0.76 \pm 1.07$ with the peak of the emission at $\lambda = 42.7$ \micron.  Photometry data points are from this work (SOFIA) and archival databases.}
\end{figure*}

\begin{figure*}
\plotone{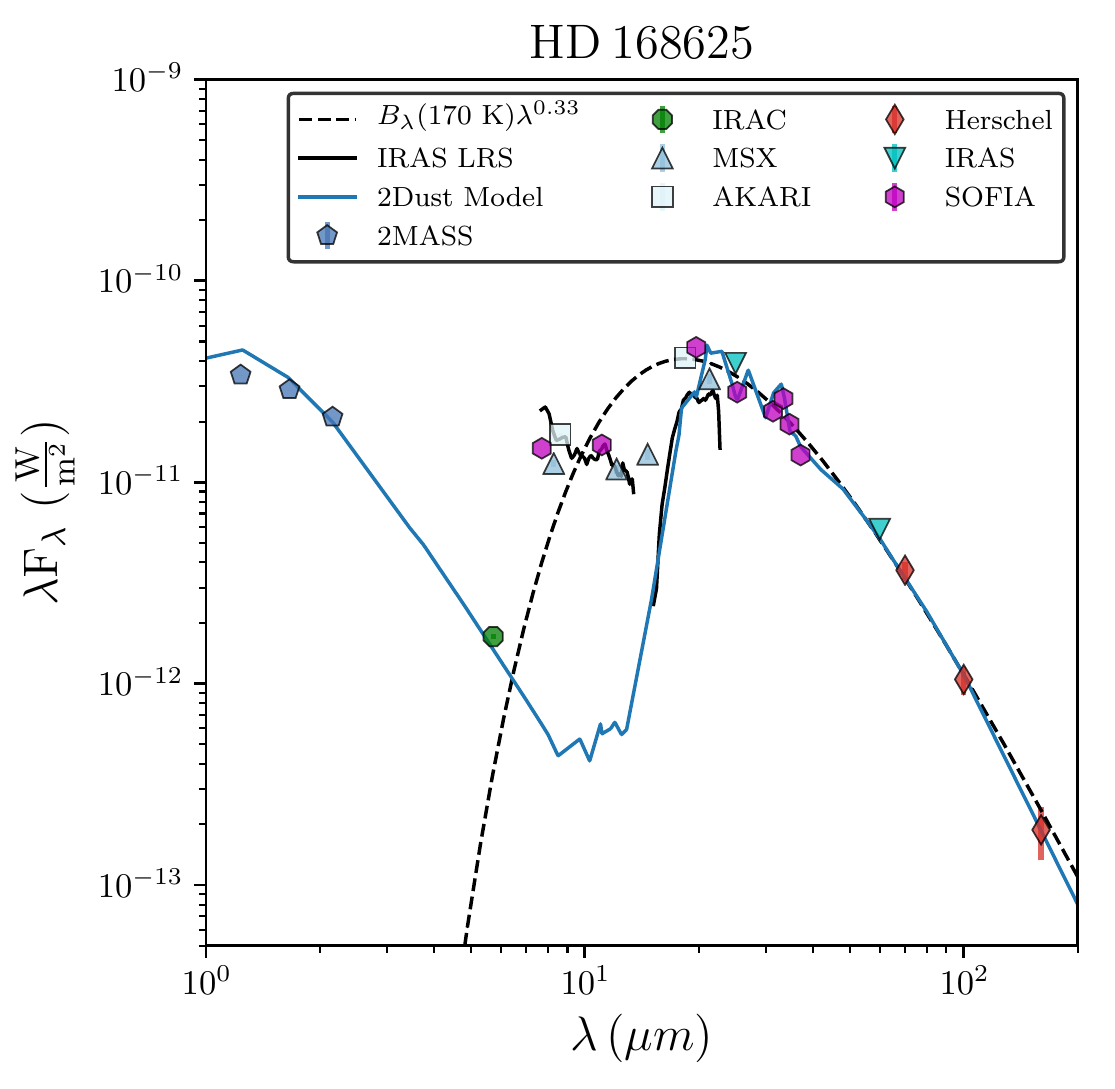}
\figcaption{\label{HD_sed}Observed and model SEDs of HD 168625.  The reddened model SED is shown as a solid blue line.  The solid black line is the \emph{IRAS} LRS \citep{Volk89}.  The dashed line is the best fitting $Q_{\lambda}\cdot B_{\lambda}(T_{d})$ function with an assumed power law emissivity $Q_{\lambda} \propto \lambda^{-\beta}$, which yields $T_{d}$ = $170 \pm 40$ K, $\beta$ = $-0.33 \pm 1.04$ with the peak of the emission at $\lambda = 18.3$ \micron.  Photometry data points are from this work (SOFIA), and archival databases.}
\end{figure*}

 We present the IR spectral energy distributions (SEDs) of MN 90 and HD 168625 in Figure \ref{MN_sed} and \ref{HD_sed}, respectively.  To supplement our newly acquired SOFIA/FORCAST photometry, we also gathered archival data from the NASA/IPAC Infrared Science Archive (IRSA; \citealp{Berriman08}) database.  These include photometry from the Two Micron All-Sky Survey (2MASS; \citealp{Skrutskie06}) at 1.25, 1.65, and 2.17 \micron, the \emph{AKARI} satellite \citep{Murakami07} at 9 and 18 \micron, the \emph{Wide-field Infrared Survey Explorer} (\emph{WISE}; \citealp{Wright10}) at 3.4, 4.6, 12 and 22 \micron, the \emph{Midcourse Space Experiment} (\emph{MSX}; \citealp{Egan03}) at 8.3, 12.1, 14.7, and 21.3 \micron\, the \emph{Spitzer} MIPS at 24 \micron, and the \emph{Spitzer} Infrared Array Camera (IRAC) Galactic Legacy Infrared Midplane Survey Extraordinaire I (GLIMPSE I; \citealp{Benjamin03, Churchwell09}) program at 3.6, 4.5, 5.8, and 8 \micron.  Photometry from our SOFIA/FORCAST images and the \emph{Herschel} \citep{Pilbratt03} Photoconductor Array Camera and Spectrometer (PACS; \citealp{Poglitsch10}) at 70, 100, and 160 \micron\ were obtained by using aperture photometry after sky background subtraction.  For MN 90 we used an aperture of 30\arcsec\ and a sky annulus with an inner radius of 31.5\arcsec\ and outer radius of 35\arcsec.  For HD 168625 we used an aperture of 20\arcsec\ and a sky annulus with an inner radius of 21.5\arcsec\ and outer radius of 25\arcsec.  Note that in the longer wavelength \emph{Herschel} PACS/SPIRE images of MN 90 from $160-500$ \micron\ the dust shell is no longer distinguishable above bright diffuse background emission.  Only IRAC 5.8 \micron\ photometry of the central star was available for HD 168625, the central star saturated the IRAC detector in the 3.6 and 4.5 \micron\ images, while warm circumstellar dust saturated the detector at 8.0 \micron.  For HD 168625 we further include the \emph{Infrared Astronomical Satellite} (\emph{IRAS}; \citealp{Neugebauer84}) Low Resolution Spectrometer (LRS) spectra from \citet{Volk89}.
 
\begin{deluxetable}{ccccc}
\tablecaption{SOFIA/FORCAST Fluxes from HD 168625 and MN 90 \label{tab:phot}}
\tablewidth{.5\textwidth}
\tabletypesize{\footnotesize}
\tablecolumns{3}
\tablehead{\colhead{Filter} & \colhead{Flux} & \colhead{Error}\\
\colhead{} & \colhead{$10^{-12} (\rm{W\, m^{-2}})$} & \colhead{$10^{-12} (\rm{W\, m^{-2}})$}}
\startdata
\multicolumn{3}{c}{MN 90}\\
\hline
F371 & 2.74 & 0.01 \\
\hline\\
\multicolumn{3}{c}{HD 168625}\\
\hline
F077 & 14.75 & 0.04  \\
F111 & 15.32 & 0.05 \\
F197 & 46.76 & 0.18  \\
F253 & 28.00 & 0.06  \\
F315 & 22.51 & 0.05  \\
F336 & 26.03 & 0.05  \\
F348 & 19.43 & 0.03  \\
F371 & 13.65 & 0.04 \\
\enddata
\end{deluxetable}

We model the dust shell flux as equilibrium thermal emission from dust grains all at a single radius $r_{d}$ from the star.  We assume the grains' emissivity behaves as a power law with $Q_{\lambda} \propto \lambda^{-\beta}$.  For MN 90, a least-squares fit of a $B_{\lambda}(T_d)$ curve modified with this emissivity yields $T_{d}$ = $59 \pm 10$ K, $\beta$ = $0.76 \pm 1.07$, with the peak of the emission at $\lambda = 42.7$ \micron.

For HD 168625, a least-squares fit of a $B_{\lambda}(T_d)$ curve modified with this emissivity $Q_{\lambda} \propto \lambda^{-\beta}$ yields $T_{d}$ = $170 \pm 40$ K, $\beta$ = $-0.33 \pm 1.04$ with the peak of the emission at $\lambda = 18.3$ \micron.

\subsection{Imaging of MN 90}
\begin{figure*}
\centering
\includegraphics[width=\textwidth]{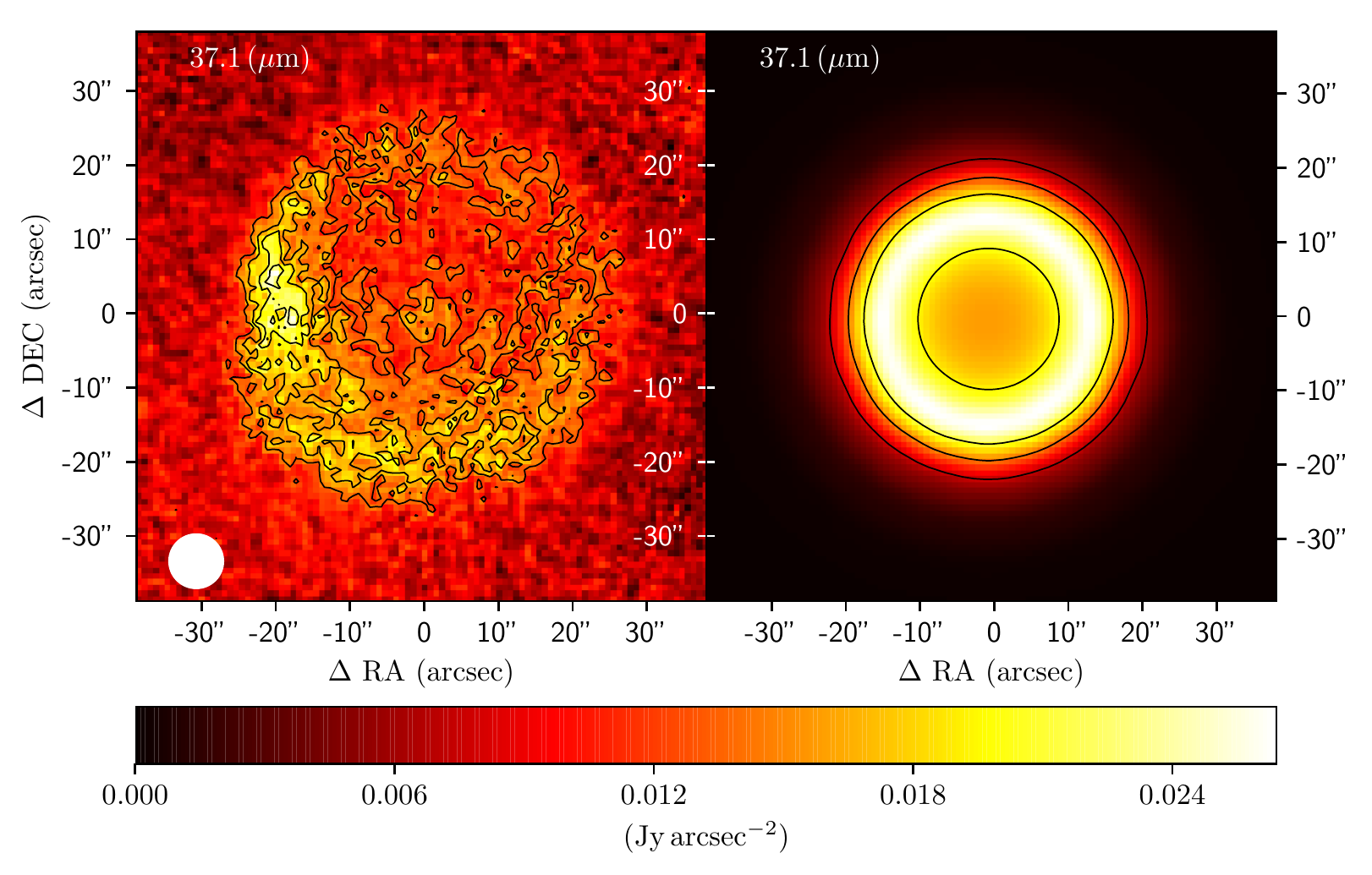}
\figcaption{\label{MN_371}Observed SOFIA/FORCAST image (left) and \twodust\ model image (right) of MN 90 at 37.1 \micron\ with north up and east to the left.  The models have been scaled to the same total flux as the observed image and convolved with a Gaussian with FWHM equal to the PSF of the SOFIA/FORCAST image (inset white circle).  In the observed SOFIA/FORCAST image, the contours are spaced at $1\sigma$, $2\sigma$, and $3\sigma$ intervals above the background noise.  In the \twodust\ model, the contours are space at 20\% intervals of the peak intensity.  The central star is not included in the model.}
\end{figure*}

Our FORCAST 37.1 \micron\ image of MN 90 is shown in Figure \ref{MN_371}.  The nebula is clearly resolved, with a radius of $\sim 15\arcsec$.  It appears nearly circular as projected on the sky, with brightened limbs enhanced towards the southeast.  The appearance is consistent with the previously obtained MIPSGAL image at 24 \micron \citep{Carey09}.  Based on its appearance we treat the nebula as a thin, hollow shell whose far-IR spectrum is due to equilibrium thermal emission from dust grains all at the same distance from the star.  This type of shell can result, for example, when a post-red supergiant (post-RSG) star has developed a fast wind during an LBV phase which sweeps up gas and dust lost during the previous RSG phase or by a previously ejected LBV shell that is swept up by the post eruption wind \citep{Smith14}.

\subsection{Imaging of HD 168625}
The FORCAST images of HD 168625 are shown in Figure \ref{HD_images}.  The nebula is clearly resolved, with a partially complete ring structure that has two peaks almost symmetric around the star.  We concur with \citet{Meixner99, OHara03} in their interpretation of these two peaks as limb-brightened peaks of a torus of dust with a radius of $\sim10\arcsec$.  The appearance is consistent with previously obtained images at 8.8, 12.5, and 20.6 \micron\ from \citet{Meixner99} and PACS 70 \micron\ images \citep{Groenewegen11}.  We stress that our SOFIA/FORCAST images do not detect the outer polar rings seen in \emph{Spitzer} IRAC images \citep{Smith07}, suggesting that the rings must be cold and below the sensitivity limits of SOFIA/FORCAST.  The emission detected with the ring morphology in the IRAC band 4 image was probably PAH emission or atomic line emission, not thermal emission from warm dust.  Figure \ref{HDTempMap} shows the temperature map that was derived from stacking the $\lambda F_{\lambda}$ SOFIA/FORCAST $7.7-37.1$ \micron\ images and performing a least-squares fit of the dust temperature, $T_d$,  using the best fit modified blackbody of the SED (i.e. $B_{\lambda}(T_{d}) \cdot \lambda^{0.33})$ at each pixel location.  The images were centered relative to one another by comparing the locations of the limb brightness peaks, and the $7.7-33.6$ \micron\ images were convolved with a 2D Gaussian kernel with a FWHM of 3.5\arcsec\ to match the resolution of the 34.8 and 37.1 \micron\ images.  Our temperature map shows a large gradient in dust temperatures with inner torus temperatures of $\sim 180$ K and outer temperatures of $\sim 80$ K, which is in agreement with the estimate of $170\pm40$ K obtained from our least-squares fit to the IR excess but is slightly higher than the equilibrium temperature estimates made by \citet[113 K]{Pasquali02}, \citet[135 K]{Robberto98b} and \citet[130 K]{OHara03}.  The inaccuracies of this temperature map are due in large part to the fact that the method used to create it assumes that emission is purely thermal and that the dust shell is in thermal equilibrium. As noted previously, much of the $8.8-12.5$ \micron\ flux arises from transient, non-equilibrium emission from PAH grains. Therefore, using images at these wavelengths to derive quantitative conclusions from the temperature map has some limitations, however, we can interpret the maps qualitatively as discussed in Section \ref{discussion}.

\subsection{IR Reddening}
\label{reddening}
The mid-IR photometry (5 -- 40 \micron) must be de-reddened for comparison with the \twodust\ model SED outputs, or conversely, the \twodust\ SEDs must be reddened, as we have done.  We used the \citet{Fritz11} extinction law as it utilizes the most recent near-IR (1 -- 2.4 \micron) observations of the galactic center.  We extend the law longward of 24 \micron\ by adopting the \citet{Draine03} interstellar extinction curve defined in Figure 10 of that paper, as was done by \citet{Lau13}.  We scaled the \AKs of the \citet{Fritz11} extinction law based on the distance to the objects.  We accomplished this by utilizing the 3-D Milky Way dust map published by \citet{Green15} to estimate the extinction.  We then converted from extinction to reddening by assuming \AKs$ = 0.320 \times \rm{E(B-V)}$, as calculated by \citet{Yuan13} for a 7000 K source spectrum at $\rm{E(B-V)} = 0.4$ mag, using the \citet{Cardelli89} reddening law and assuming $\rm{R_V} = 3.1$.  

As mentioned in Section \ref{intro}, the distance to MN 90 is unknown.  In the absence of further information about the star beyond the catalogued $1.2$ $-$ 8 \micron\ photometry, we consider the implications of assuming that MN 90 is an LBV.  We assume a luminosity of $L_{\star} = 3\times10^{5}$ $L_{\sun}$, at the lower end of luminosities for LBVs in their quiescent state between outbursts (see e.g. Figure 1 of \citealp{Smith04}).  This luminosity corresponds to $T_{\star} \approx$ 14000 K.  For dust grains with emissivity $Q_{\lambda} \propto \lambda^{-\beta}$, the radius $r_{d}$ of the shell may be computed with:
\begin{equation}
r_{d}^{2} = \frac{L_{\star}}{16\pi\sigma T_{d}^{4}} \left( \frac{T_{\star}}{T_{d}}\right)^{\beta}
\end{equation}
where $T_{d}$ is the grains' equilibrium temperature.  Substituting the assumed values for the star and the fitted $T_{d}$ $=$ 59 K obtained using $\beta = 0.76$ yields $r_{d}$ = 0.47 pc.  For the shell's observed angular radius of 20$\arcsec$ this places MN 90 at a distance of 4.8 kpc.  At this distance we estimate a reddening of \AKs$ = 0.52$ for MN 90.

The \emph{Gaia} DR2 \citep{Brown18} distance to HD 168625 is $1.61\pm0.17$ kpc or 1.55 kpc using the Bayesian-inferred distance from \citep{BailerJones18}.  We adopt a distance of 1.55 kpc for HD 168625 and estimate a reddening of \AKs$ = 0.32 $ for HD 168625.  Note that any visual wavelength values are subject to considerable uncertainty, as \citet{Fritz11} point out that there are several possible extrapolations from their anchor region around Brackett-$\gamma$ (2.166 \micron) into the visual. 

\section{Radiative Transfer Modeling}
\label{model}
\subsection{\twodust\ Introduction}
We utilize the the axisymmetric radiative transfer code \twodust\ \citep{Ueta03} to estimate the dust mass and dust shell morphology of MN 90 and HD 168625.  The code solves the equation of radiative transfer following the principle of long characteristic (i.e. traces the radiation hitting the dust grain from anywhere in the shell including the star and other dust radiation) in a 2-D polar grid, while considering a 3-D radiation field at each grid point.  The dust opacities are calculated using Mie theory from a user-given size distribution and optical constants of the dust grains.  It can be used to model a variety of axisymmetric astronomical dust systems.  The dust distribution is expressed analytically as 

\begin{fleqn}[0pt]
\begin{align}
&\resizebox{0.47\textwidth}{!}{$\rho(r,\theta) = \rho_{min}\Big(\frac{r}{r_{min}}\Big)^{-B\big(1+C\sin^F\theta\{\exp[-(r/r_{sw})^D]/\exp[-(r_{min}/r_{sw})^D]\}\big)}$} \\ \nonumber
&~~~\times \big[1 + A(1 - \cos\theta)^F \\ \nonumber
&~~~\times \{\exp[-(r/r_{sw})^E]/\exp[-(r_{min}/r_{sw})^E]\}\big]\nonumber
\end{align}
\end{fleqn}
\noindent
where $r$ is the radius within the limits of $r_{min}$ and $r_{max}$, $r_{sw}$ is the boundary between the spherical
AGB wind and the axisymmetric superwind, $\theta$ is the latitude, and $\rho_{min}$ is the dust mass density on the equatorial axis at the inner edge of the envelope.  The letters $A-F$ are input parameters that define the geometry of the dust density profile.  A, changes the overall axisymmetric structure to the shell, which can be made disk-like or toroidal by the parameter F.  The parameter B determines the radial fall-off of the profile and can be a function of the latitudinal angle, $\theta$, through the parameter C.  These parameters determine the toroidal structure of the innermost region of the shell, which is considered to be caused by an axisymmetric superwind at the end of the AGB phase.  The mid-region of the shell assumes a somewhat spheroidal dust distribution reflecting the transition of mass loss geometry from spherical to axial symmetry during the course of the AGB mass loss history. The parameters D and E control the abruptness of the transition in the shell: small values correspond to a slow transition and large values correspond to an abrupt transition.  We used a \citet{Mathis77} power law grain size distribution:
\begin{equation}
n(a) = a^{-3.5}  ~~,~~  a_{min} < a < a_{max}
\end{equation}
where $a_{min}$ is the minimum grain size and $a_{max}$ is the maximum grain sized, as specified by inputs.  Because \twodust\ is axisymmetric, it is not possible to create dust shells with different size parameters in different lobes of the nebula.  A more extensive discussion of the geometric parameters given in Equation (2) is given in \citet{Ueta03}.  For more examples of the use of \twodust, see \citet{Ueta01a, Ueta01b, Meixner02, OHara03, Meixner04}.

\subsection{Input Parameters}
\label{input_params}
\citet{OHara03} previously analyzed the morphology and parameters for the circumstellar dust around HD 168625 using \twodust, therefore, we adopt their parameters as initial values for our model.  

Previous mid-IR spectra of HD 168625 by \citet{Skinner97} indicate the presence of silicates in the dust shell of HD 168625 and previous studies of the dust shells surrounding the LBVs Wra 751 and AG Car indicate that amorphous silicates are the dominant species \citep{Voors00}.  Observations of HD 168625 by \citet{Volk89, Skinner97, Umana10} have identified polycyclic aromatic hydrocarbons (PAHs), which is evident in the \emph{IRAS} LRS spectra and suggests the possible presence of carbonaceous grains as well.  There is also evidence of crystalline forsterite grains being present \citep{Blommaert14}, which may contribute to the emission seen at $\sim$ 11 and 19 \micron\  in the \emph{IRAS} LRS.  Therefore, we considered a complex dust distribution model composed of amorphous olivine (MgFeSiO$_4$; \citealp{Dorschner95}), crystalline forsterite (Mg$_2$SiO$_4$; \citealp{Servoin73, Scott96}), and amorphous carbon \citep{Rouleau91}.  However, because \twodust\ cannot account for transiently heated very small dust grains or PAH emission, we ignore fitting the \twodust\ model to the 8 -- 15 \micron\ region.

Most estimates of the effective temperature of the central star of HD 168625 are between 12000 -- 15000 K \citep{Nota96, OHara03, Mahy16}, and may vary by a few thousand degrees.  The temperature and radius of the star have been adjusted to roughly match the observed SED.  However, since the data points are taken by many observers over several decades, we cannot model HD 168625 at any single epoch.  The inner radius of the dust shell is well constrained by these mid-IR images to be 8.5\arcsec\ or 0.06 pc at a distance of 1.55 kpc.

As previously mentioned, little is know about MN90 and the central star.  We use the SOFIA/FORCAST 37.1 \micron\ image to constrain the angular size of $r_{min}$ to a value of 20\arcsec\ or 0.47 pc at a distance of 4.8 kpc.  We started with a stellar temperature of 14000 K and radius 100 $R_\sun$.  Like HD 168625, the temperature and radius of the star have been adjusted to roughly match the observed SED and we used a single species dust distribution of amorphous silicates.

\subsection{Model Results}
\label{model_results}
We ran approximately 150 models to obtain the best fit to the SED and SOFIA/FORCAST images of both stars.  When available, stellar and dust parameters were taken from previous observations in the literature.  We started our modeling by fitting the stellar parameters $T_*$ and $R_*$.  The mid-IR images constrain the inclination angle ($\theta_{inc}$), inner radius of the dust shell ($r_{min}$), and dust density function parameters A--F.  We then fit the mass fraction of the mineral species, minimum ($a_{min}$) and maximum grain sizes ($a_{max}$), and optical depth at 37.1 \micron\ ($\rm \tau_{37.1\micron}$).  We constrained the maximum and minimum grain sizes to be between $0.001-2.0$ \micron .  The best fit model was determined by eye.  The best fit SEDs of MN 90 and HD 168625 are shown in Figures \ref{MN_sed} and \ref{HD_sed}.  Input dust distribution parameters for both stars are given in Table \ref{tab:dust_params}.  Input and derived stellar and dust parameters for the best fit models of MN 90 are given in Table \ref{tab:MN_params}.   Input and derived stellar parameters for the best fit model of HD 168625 are given in Table \ref{tab:HD_params} with the dust parameters summarized in Table \ref{tab:HD_dust}.

\begin{figure*} 
\gridline{\fig{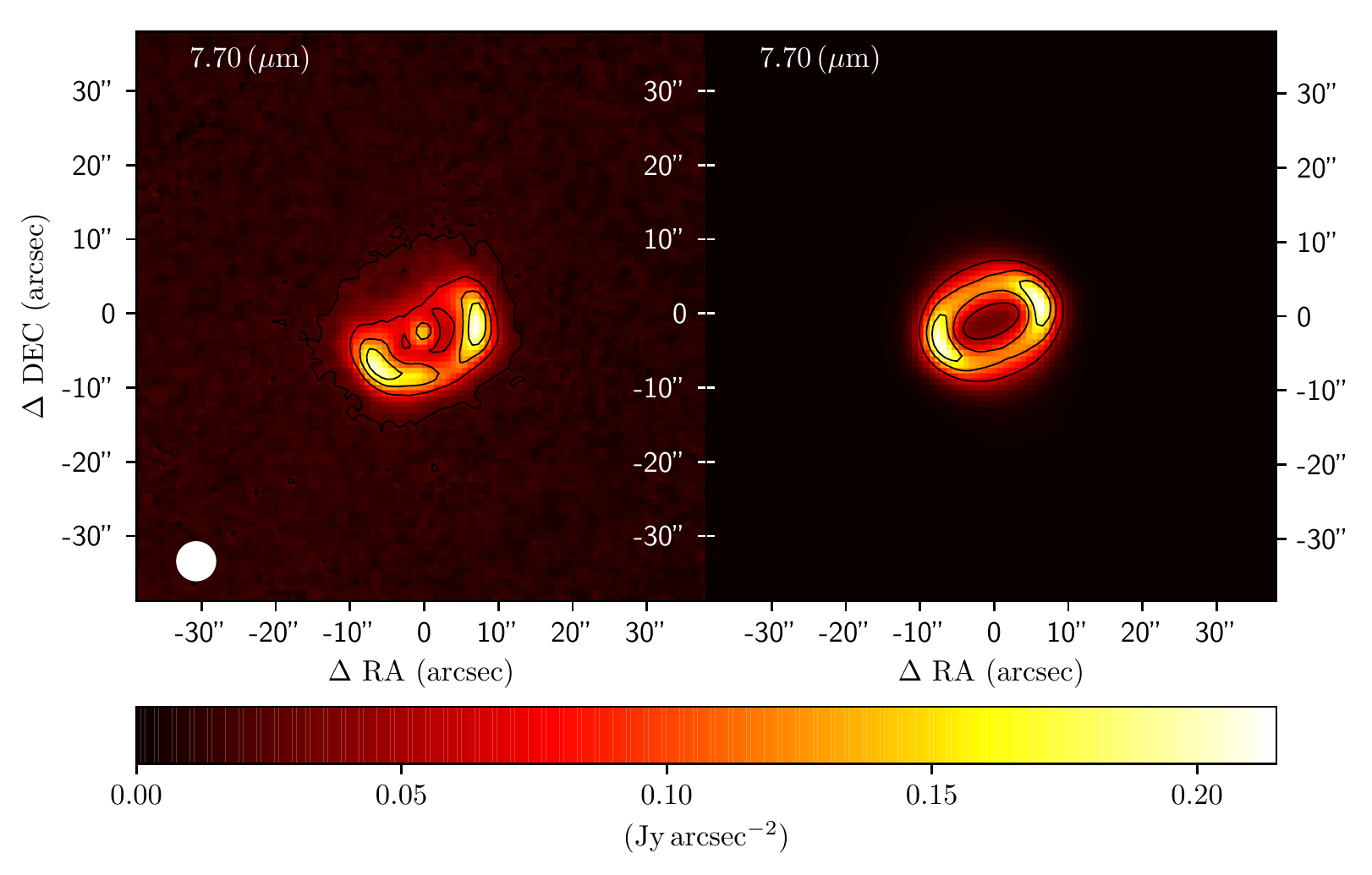}{0.8\textwidth}{(a) FORCAST F077}}
\gridline{\fig{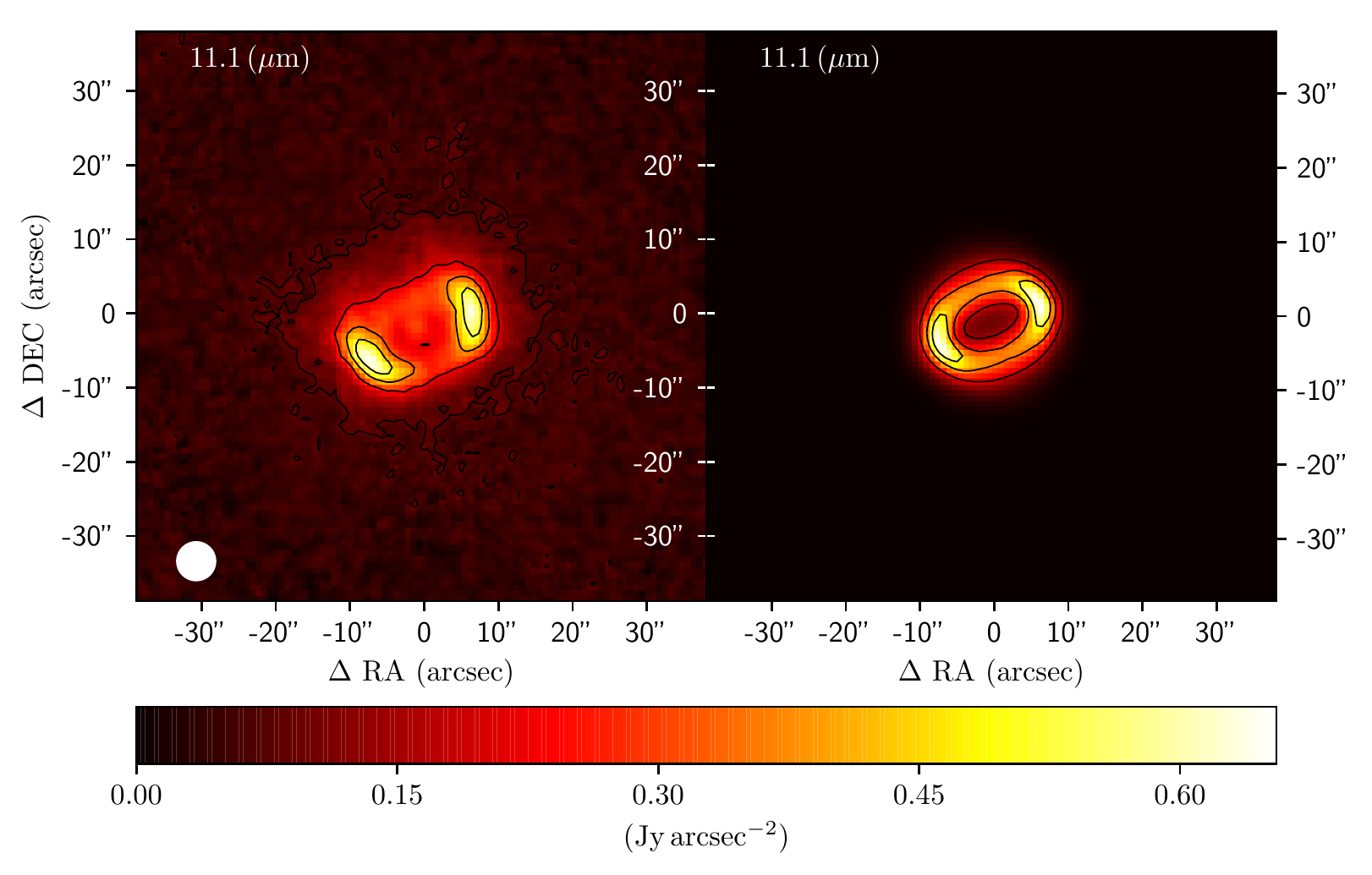}{0.8\textwidth}{(b) FORCAST F111}}
\end{figure*}
\begin{figure*}
\gridline{\fig{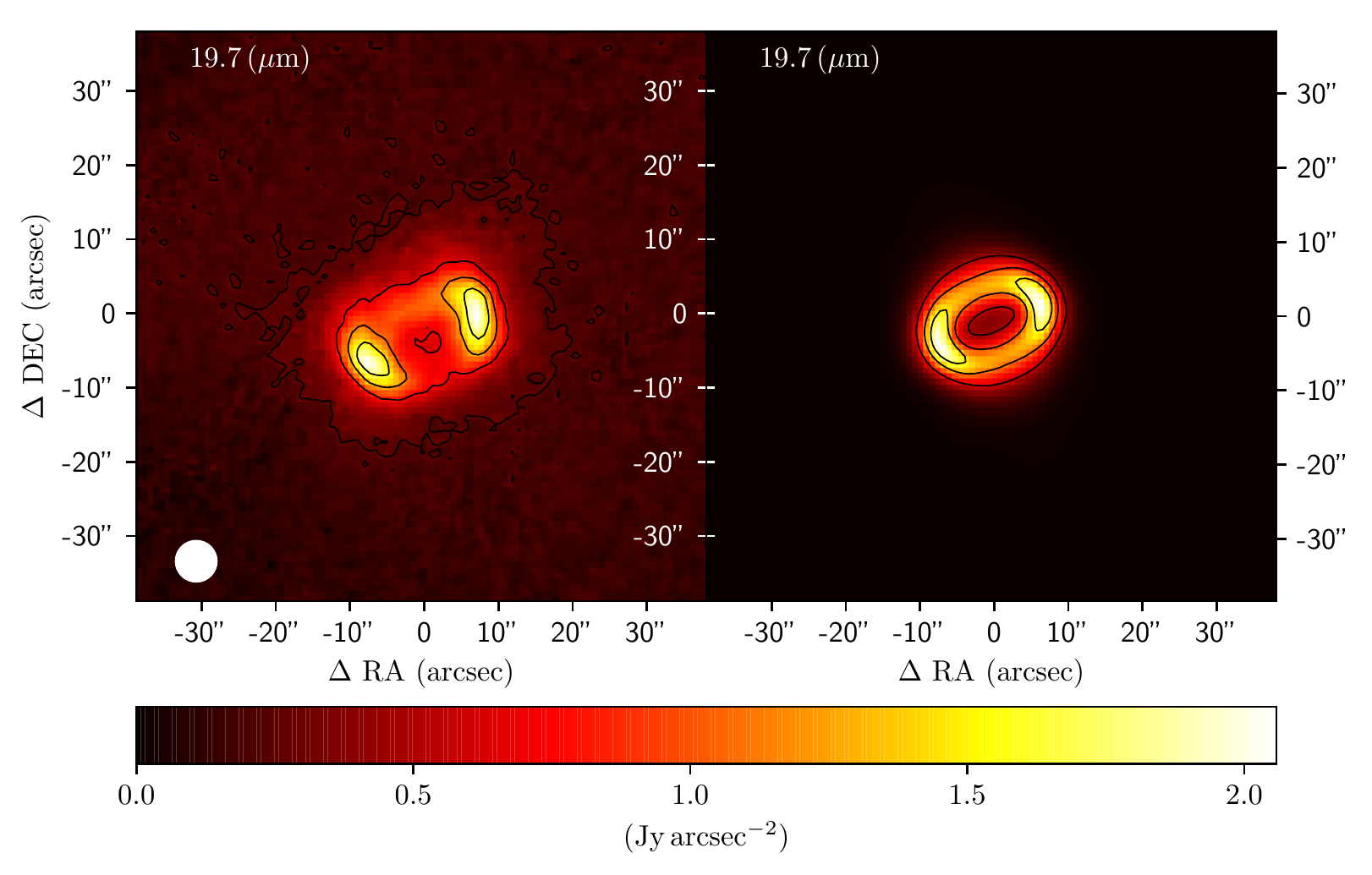}{0.8\textwidth}{(c) FORCAST F197}}
\gridline{\fig{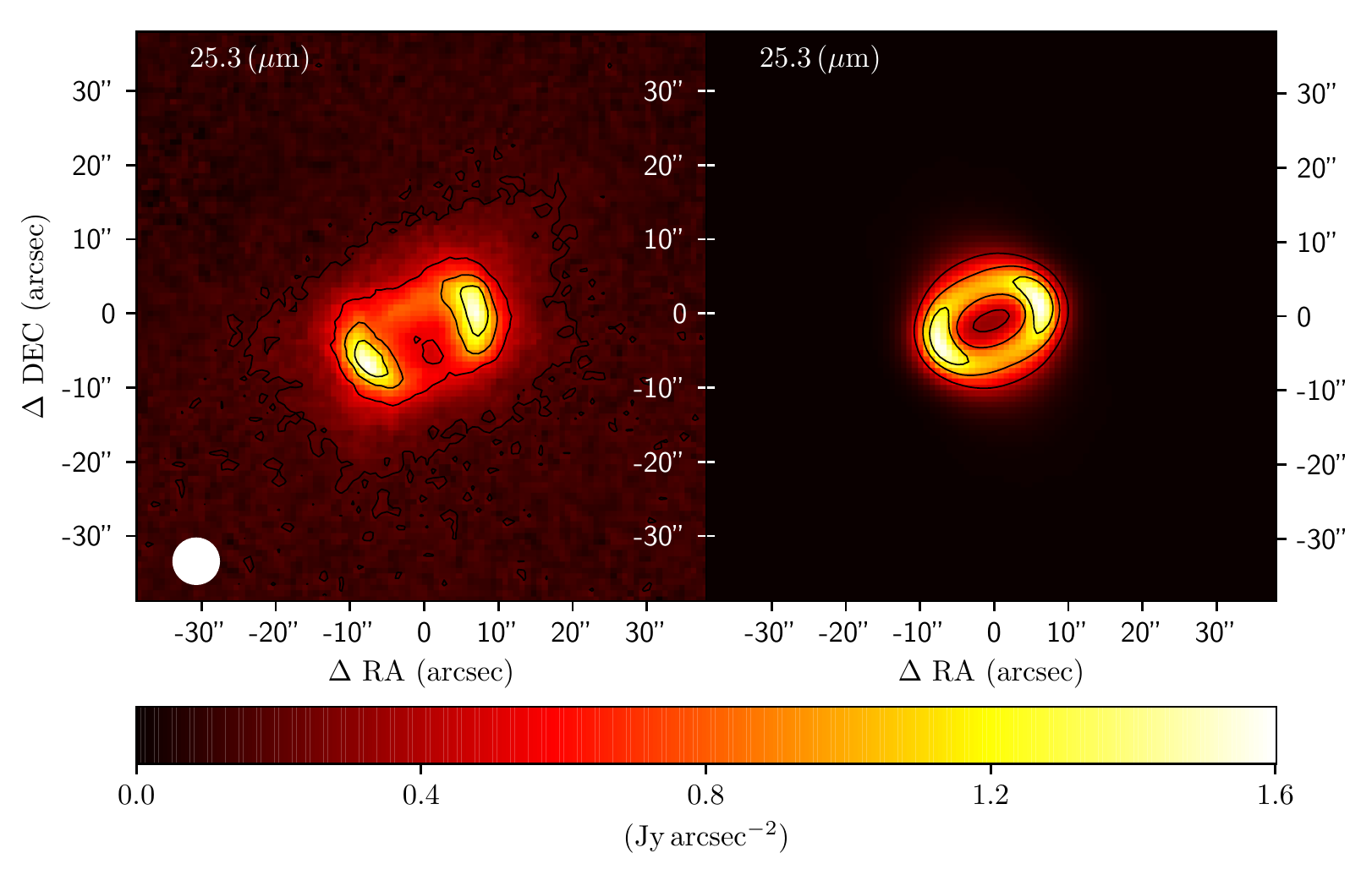}{0.8\textwidth}{(d) FORCAST F253}}
\end{figure*}
\begin{figure*}
\gridline{\fig{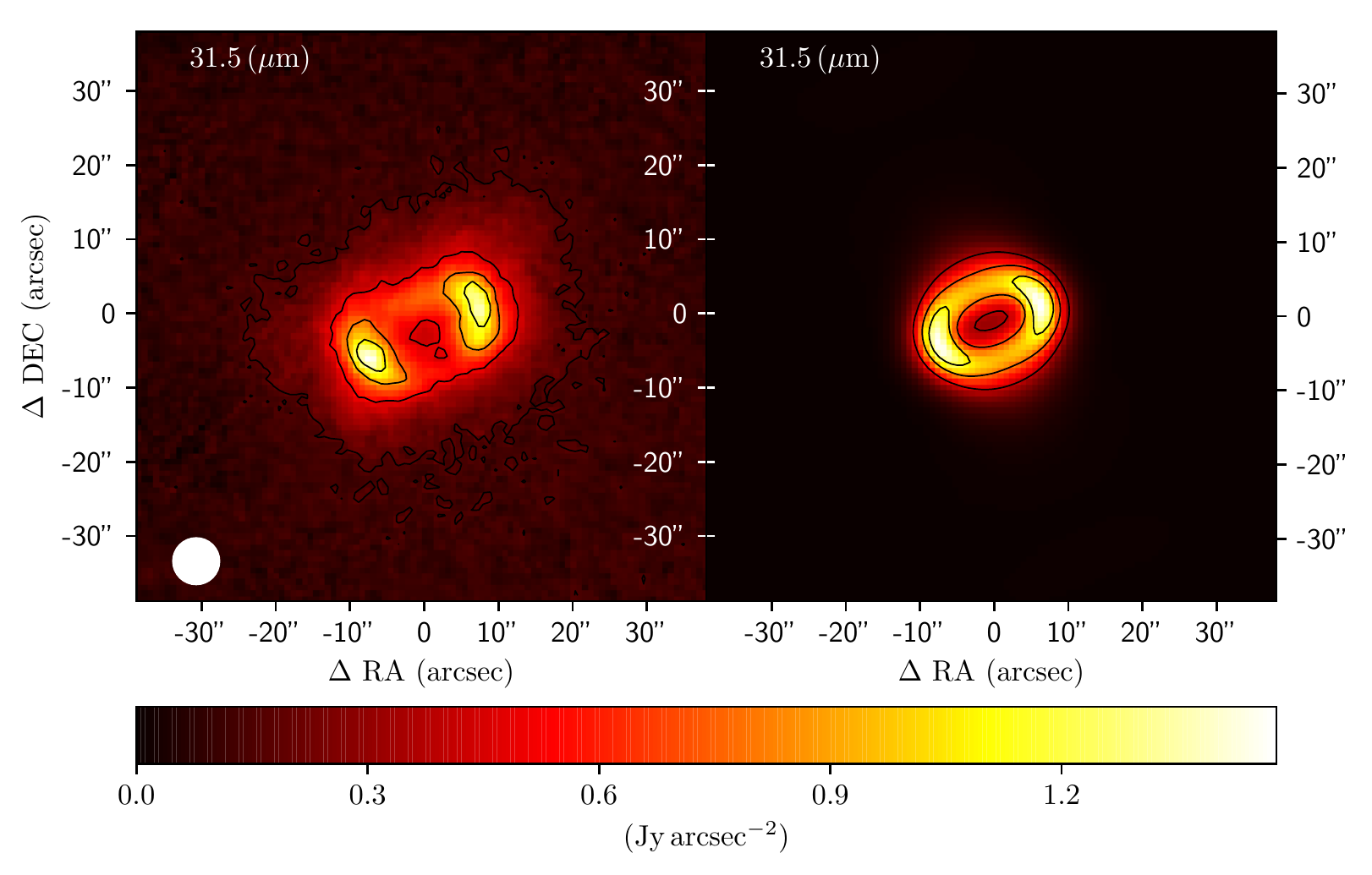}{0.8\textwidth}{(e) FORCAST F315}}
\gridline{\fig{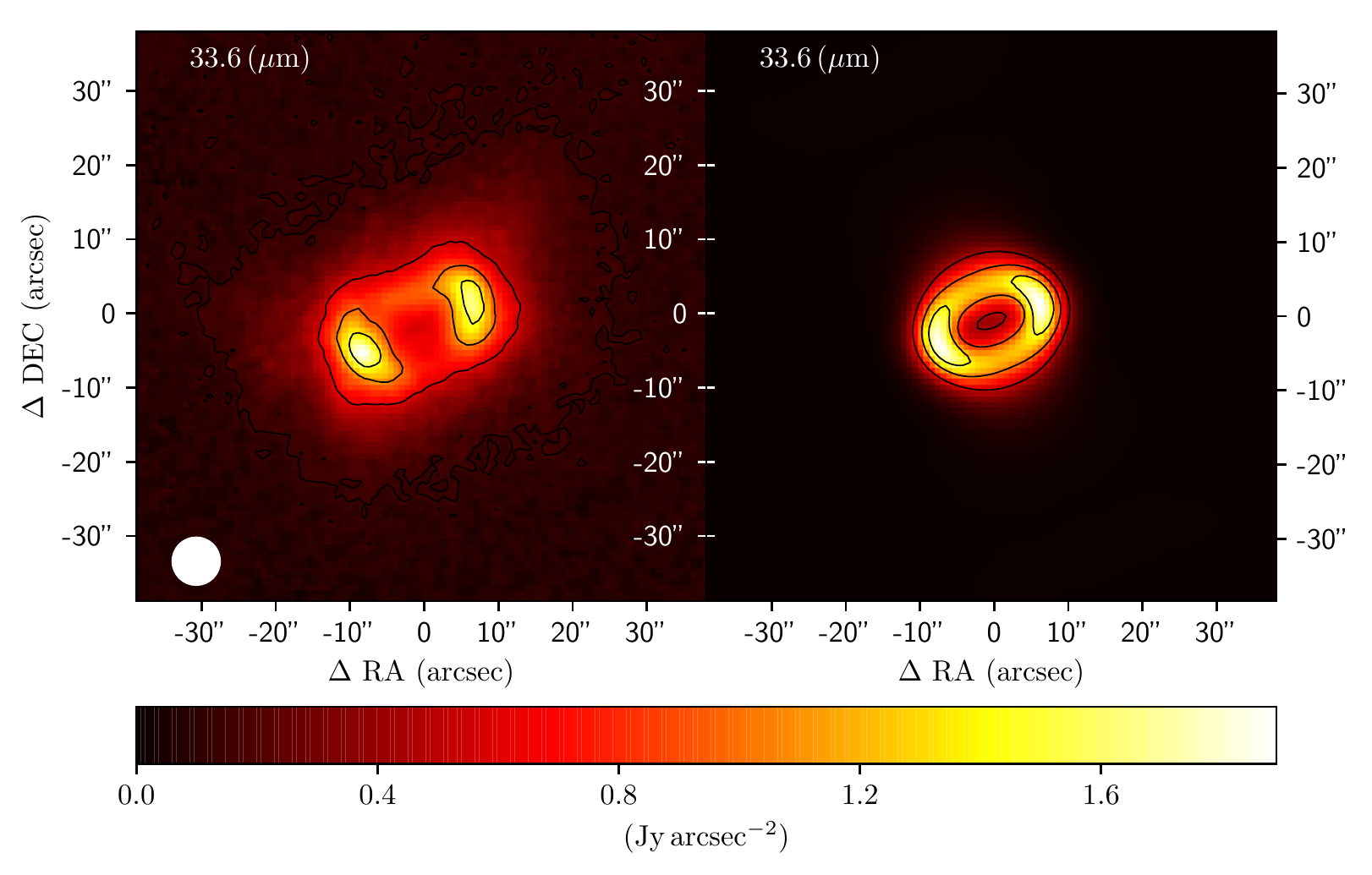}{0.8\textwidth}{(f) FORCAST F336}}
\end{figure*}
\begin{figure*}
\gridline{\fig{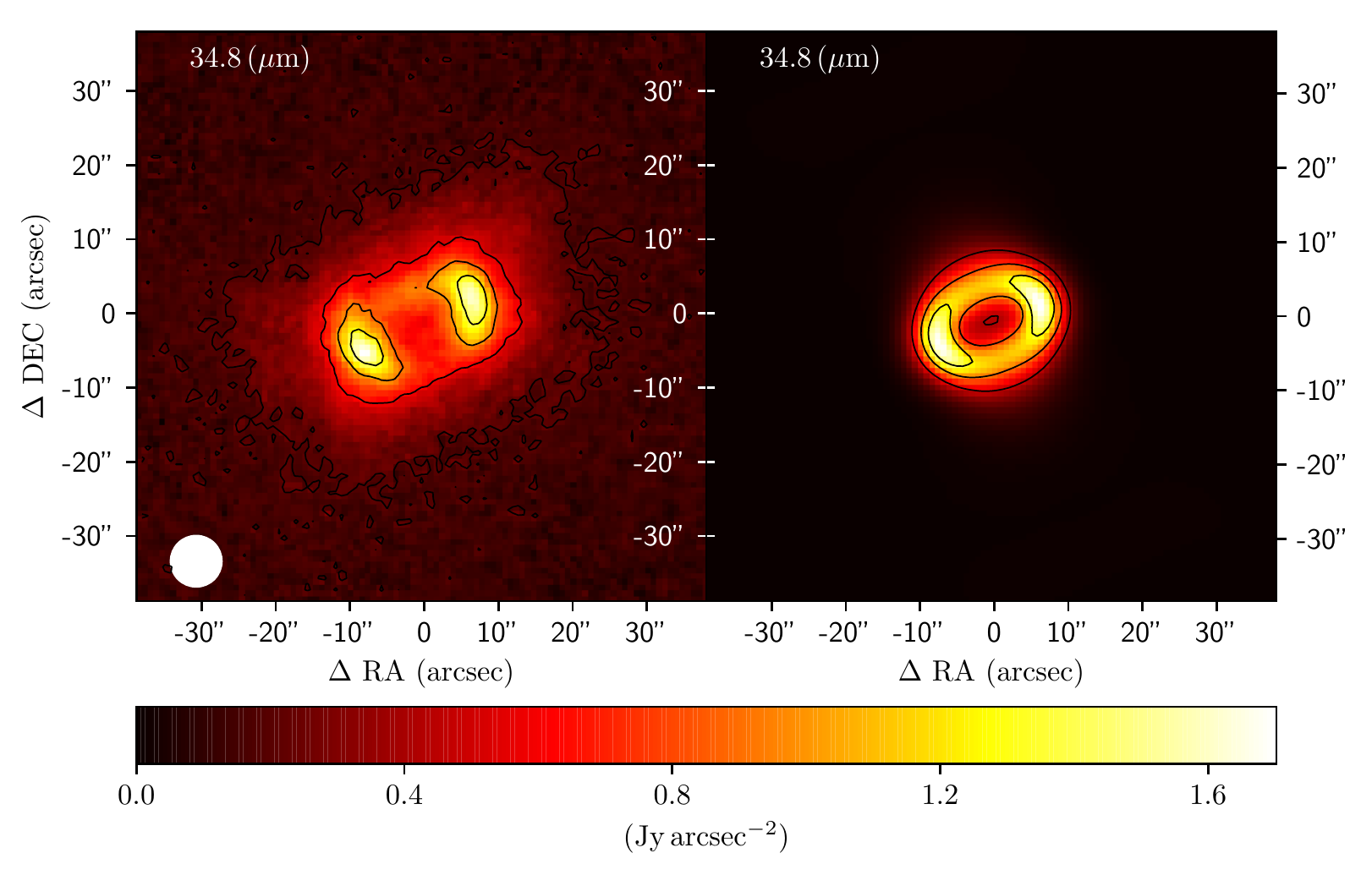}{0.8\textwidth}{(g) FORCAST F348}}
\gridline{\fig{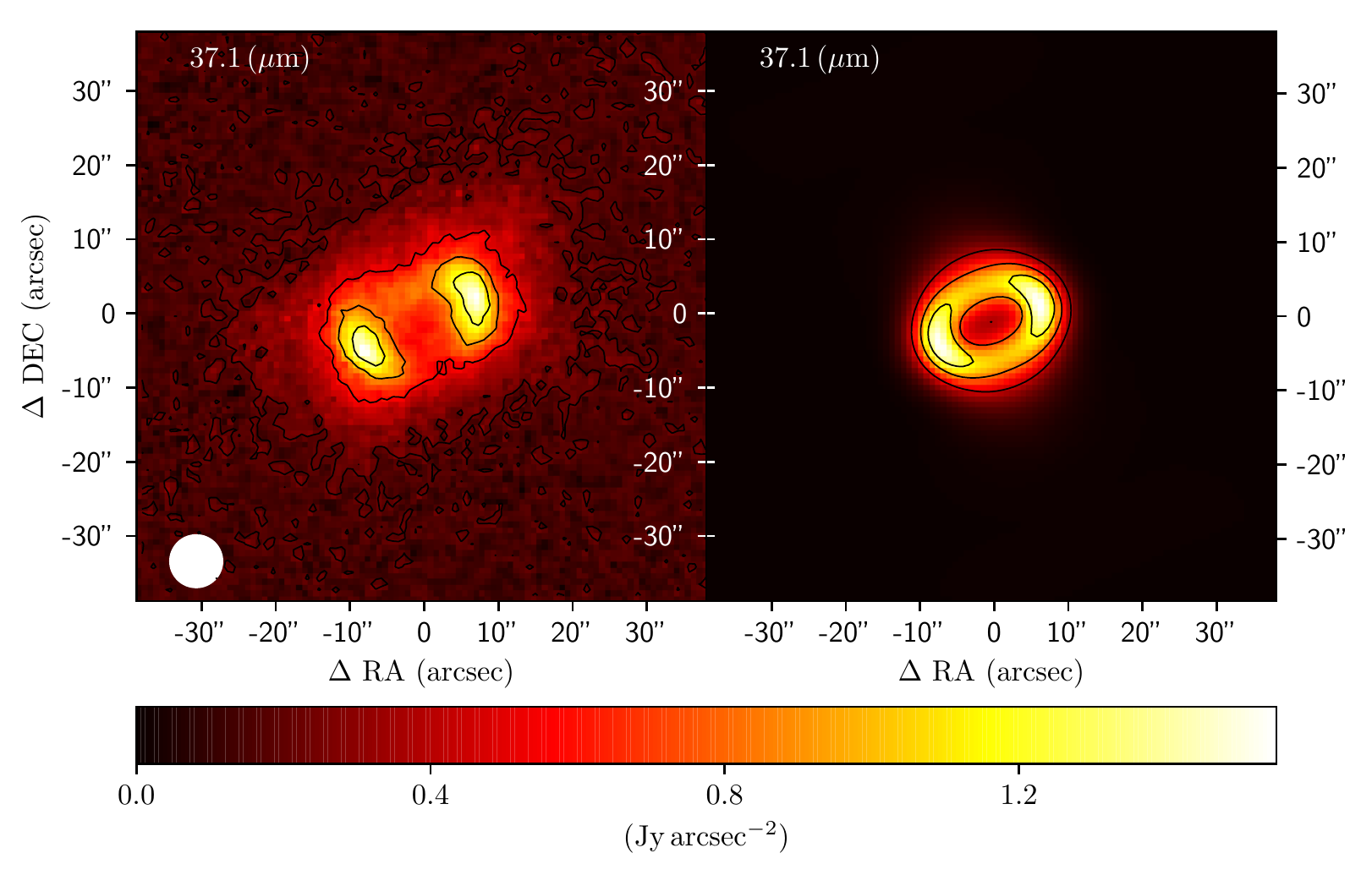}{0.8\textwidth}{(h) FORCAST F371}}
\figcaption{\label{HD_images}Observed SOFIA/FORCAST image (left) and \twodust\ model image (right) of HD 168625 at (a) 7.7 \micron, (b) 11.1 \micron, (c) 19.7 \micron, (d) 25.3 \micron, (e) 31.5 \micron, (f) 33.6 \micron, (g) 34.8 \micron, and (h) 37.1 \micron\ with north up and east to the left. The models have been scaled to the same total flux as the observed image and convolved with a Gaussian with FWHM equal to the matching PSF of the SOFIA/FORCAST image (inset white circle).  In the observed SOFIA/FORCAST images, the contours are spaced at $1\sigma$, $2\sigma$, and $3\sigma$ intervals above the background noise.   In the \twodust\ model, the contours are space at 20\% intervals of the peak intensity.  The central star is not included in the model.}
\end{figure*}

\begin{figure*}
\centering
\includegraphics[width=\textwidth]{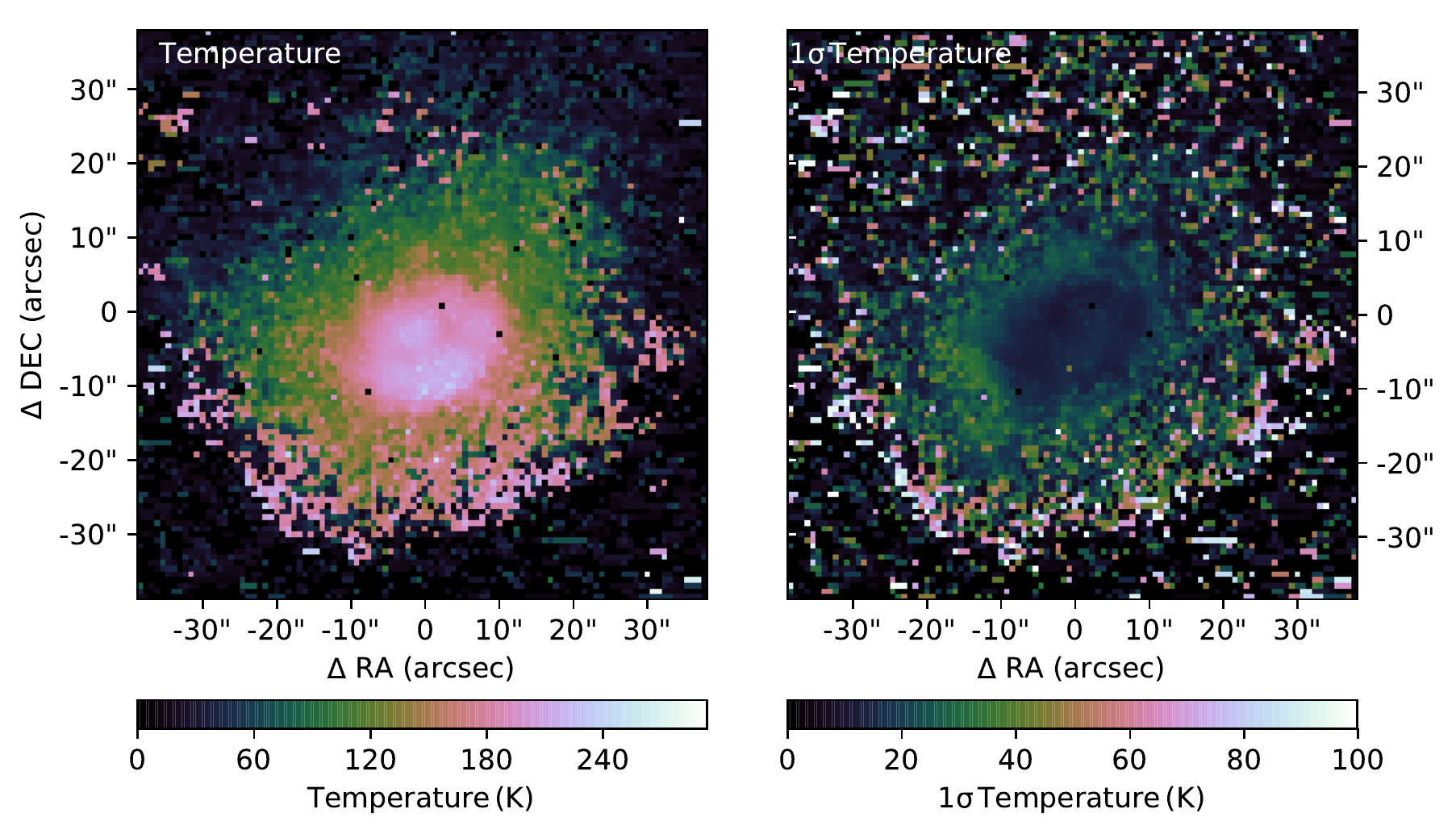}
\figcaption{\label{HDTempMap}Temperature map (left) and 1$\sigma$ temperature error map (right) of HD 168625, derived from stacking the $\lambda F_{\lambda}$ SOFIA/FORCAST 7.7--37.1 \micron\ images and performing a least-squares fit of the dust temperature, $T_d$,  using the best fit modified blackbody of the SED (i.e. $B_{\lambda}(T_{d}) \cdot \lambda^{0.33})$ at each pixel location.  Pixels with errors larger than 100 K have been excluded.}
\end{figure*}

\begin{deluxetable}{ccc}
\scriptsize
\tablewidth{.5\textwidth}
\tablecaption{\label{tab:dust_params} \twodust\ Density Function Parameters for the Best-Fit Models}
\tablehead{
\colhead{Parameter} &  \colhead{MN~90} & \colhead{HD~168625}
}
\startdata
$A$ & 1 & 30 \\
$B$ & 2 & 2 \\
$C$ & 2.5 & 4 \\
$D$ & 0 & 6 \\
$E$ & 0 & 3 \\
$F$ & 0 & 4 \\
$\theta_{inc}$ & {$0^\circ$} & {$55^\circ$}\\
$\rm \tau_{37.1\micron}$ at eq. & $6\times10^{-4}$  & $3.3\times10^{-3}$  \\
\enddata
\end{deluxetable}

The model SED for MN 90 fits the data well in the wavelength region $\lambda \geq 8$ \micron.  The discrepancy at shorter wavelengths may be due to the choice of the anchor region for the reddening as discussed in Section \ref{reddening}, an underestimate in the distance, or incorrect stellar parameters.  Our model image of MN 90 compares well with the FORCAST 37.1 \micron\ image except for the northwest edge of the shell which appears brighter in the FORCAST image suggesting that the shell is slightly asymmetric.  Overall, however, the morphology of the dust shell is well modeled as a symmetric sphere.  We found the best fit density distribution had an amorphous silicate shell extending from $r_{min} = 0.31$ pc to $3r_{min} = 0.92$ pc.  The grain size distribution suggests that the silicates have grain sizes between 0.1 -- 0.5 \micron.  Our model gives a dust mass of $(3.2\pm0.1) \times 10^{-2}~$\solarmass\ and a dust mass-loss rate of $(7.3\pm0.4) \times 10^{-6}~$\massrate $\times\ (v_{exp}/50\,$ \kms\!).

\begin{deluxetable}{ccc}
\tabletypesize{\scriptsize}
\tablecolumns{2} 
\tablewidth{.5\textwidth} 
\tablecaption{\label{tab:MN_params}MN 90 \twodust\ Input and Derived Parameters}
\tablehead{\multicolumn{2}{c}{\phantom{\LARGE I}Input\phantom{\LARGE I}}}
\startdata 
$L_* ~\propto~ d$ & {$9.3 \times 10^{4}~L_{\sun}$} \\
$T_*$  & {$16000$ K} \\
$R_* ~\propto~ d$ & {$85~R_\sun$} \\
$d$      & {$4.8$ kpc} \\
ISM \AKs     & {$0.52$}   \\
$r_{min} ~\propto~ d$  & 0.35 pc\\
$r_{max} ~\propto~ d$ & 1.05  pc\\
$r_{sw} ~\propto~ d$ & 0.52 pc\\
$a_{min}$ & 0.1 \micron \\
$a_{max}$ & 0.5 \micron \\
$v_{exp}\tablenotemark{a}$ & {50 \kms}\\
Amorphous Silicates\tablenotemark{b} & 100\%\\
\cutinhead{Derived}
$M_{dust} ~\propto~ d^2$ & {$(3.2\pm0.1) \times 10^{-2}~M_{\sun}$}\\
$\dot M_{dust} ~\propto~ d^2$ & {$(7.3\pm0.4) \times 10^{-6}$~\massrate} \\
$\tau_{AGB}\tablenotemark{c} ~~\propto~ d$ &  {$1.0 \times 10^{4}~yr$}\\
\enddata
\tablenotetext{a}{Typical for LBVs \citep{Nota95}}
\tablenotetext{b}{MgFeSiO$_4$ \citep{Dorschner95}}
\tablenotetext{c}{Timescale for mass-loss on the AGB}
\end{deluxetable}

The model SED for HD 168625 fits the data well in most of the wavelength range except for 8 -- 15 \micron\ as discussed in Section \ref{input_params}.   We confirm the dust distribution model of \citet{OHara03} with only minor differences in the stellar temperature, radius and grain size distribution.  The morphology of the dust is found to be a torus with an equator-to-pole density ratio of 30, an elliptical midshell, and an inclination angle of $55^\circ$ with the plane of the sky.  Our model images compare well with the FORCAST observations except for the gap in the northern rim of the 7.7 \micron\ emission and the gap in the southern rim of the 19.7, 25.3, 31.5, and 33.6 \micron, emission which may indicate a different composition, azimuthal asymmetry in the dust density distribution around the ring, or the lack of large dust grains in this region.  This supports the suggestion that the dust shell may include small, transiently heated dust grains that are not at thermal equilibrium and are not accounted for in these models.  The grain size distribution suggests that grain sizes between 0.001 -- 1.0 \micron\ exist in the circumstellar environment.  Our model gives a dust mass of $(2.5\pm0.1) \times 10^{-3}~$\solarmass and a dust mass-loss rate of $(3.2\pm0.2) \times 10^{-7}~$\massrate.  If we rescale the dust mass estimates of \citet{OHara03} using the \emph{Gaia} DR2 distance, their dust mass changes to $(7.7\pm0.3) \times 10^{-4}~$\solarmass at a mass-loss rate of $(6.7\pm0.3) \times 10^{-8}~$\massrate.  The discrepancy can be attributed to both our larger grain size distribution and complex dust distribution model which results in a larger optical depth in our model and therefore a larger dust mass estimate.  Our estimate is likely an underestimate of the mass because we are only looking at material in the equator, $\sim$ 10 -- 20\% of the solid angle of the entire nebula.  The bipolar nebula seen in \emph{Spitzer} images suggests that the circumstellar material at higher latitudes was ejected with higher speed, and therefore it is farther from the star and cooler, which went undetected in the SOFIA imaging that only detects the inner torus.  The model suggests that HD 168625 lost mass in a torus-shaped outburst, which has been suggested for all LBVs by \citet{Hutsemekers94}.  The current fast wind of the LBV has probably interacted with this torus, creating an elliptical bubble perpendicular to the plane of the torus, which is consistent with a unified model of LBV nebulae proposed by \citet{Nota95}.

\begin{deluxetable}{cc}
\tabletypesize{\scriptsize}
\tablecolumns{2} 
\tablewidth{.5\textwidth} 
\tablecaption{\label{tab:HD_params}HD 168625 \twodust\ Input and Derived Parameters} 
\tablehead{\multicolumn{2}{c}{\phantom{\LARGE I}Input\phantom{\LARGE I}}}
\startdata 
$L_* ~\propto~ d$ & $4.5 \times 10^{4}~L_{\sun}$ \\
$T_*$  & $14500$ K \\
$R_* ~\propto~ d$ & $65~R_\sun$ \\
$d$      & $1.55$ kpc \\
ISM \AKs     & $0.32$   \\
$r_{min} ~\propto~ d $ & 0.06 pc\\
$r_{max} ~\propto~ d$ & 0.23 pc\\
$r_{sw} ~\propto~ d$ & 0.20 pc\\
$v_{exp}$\tablenotemark{a} & 19 \kms \\
\cutinhead{Derived}
$M_{dust} ~\propto~ d^2$ & $(2.5\pm0.1) \times 10^{-3}~M_{\sun}$\\
$\dot M_{dust} ~\propto~ d^2$ & $(3.2\pm0.2) \times 10^{-7}$ \massrate \\
$\tau_{AGB}\tablenotemark{b} ~~\propto~ d$ & $1.5 \times 10^{3}~yr$\\
\enddata
\tablenotetext{a}{\citet{Pasquali02}}
\tablenotetext{b}{Timescale for mass-loss on the AGB}
\end{deluxetable}

\begin{deluxetable*}{ccccccc}
\tabletypesize{\scriptsize}
\tablecaption{Dust Properties for Best-Fit \twodust\ Model of HD 168625 \label{tab:HD_dust}}
\tablecolumns{7}
\tablehead{\colhead{Dust Species} & \colhead{Composition} & \colhead{Structure} & \colhead{Mass Fraction} & \colhead{Density} & \colhead{Grain Size} & \colhead{Reference}} 
\startdata
Olivine & $\rm{MgFeSiO_4}$ & Amorphous & 60\% & 3.71 & $0.001-1.0$ & \citet{Dorschner95}\\
Forsterite &  $\rm{Mg_2SiO_4}$ &  Crystalline & 25\% & 3.22 &  $0.500-2.0$ &  \citet{Servoin73,Scott96}\\
Carbon &  BE sample &  Amorphous &  15\% & 1.44 &  $0.001-1.0$ &  \citet{Rouleau91}\\
\enddata
\end{deluxetable*}

\section{Discussion}
\label{discussion}
As previously mentioned in Section \ref{input_params}, crystalline forsterite has been detected around HD 168625.  This indicates that at least some of the circumstellar material, probably that confined to the torus, has undergone annealing and suggests that the circumstellar environment is similar to that of lower mass progenitors i.e. proto-planetary nebulae (PPNe).  We should point out that our model for HD 168625 included moderately sized grains (0.1 -- 1.0 \micron), slightly larger than the model found by \citet[0.001 -- 1.0 \micron]{OHara03} which points to the nebula being relatively young and unprocessed.  Aside from the similarities in morphology between PPN and LBVs, a key difference may be the duration of mass-loss.  Far-IR images at 55 \micron\ of HD 168625 by \citet{OHara03} suggests that the outer dust shell is no more than 5 times the inner radius.  Whereas, for the Egg Nebula, a well-studied PPN, 180 \micron\ images suggest that the outer shell is a few hundred times larger than the inner radius \citep{Speck00}.  The mass-loss shells of LBVs appear to be more compressed than for PPNs, suggesting that the mass-loss occurred in a more short-lived ($< 10^4$ yr) phase compared with PPNs ($\sim 10^5$ yr).
\vspace{1cm}
\subsection{MN 90}
The morphology of MN 90 lacks a large equator-to-pole mass distribution like most LBVs and is nearly spherical, similar to V4998 Sgr  \citep{Lau14}.  The total mass (gas plus dust) lost by the star and the rate of mass-loss are estimated to be about $3.2\pm0.1$ \solarmass and $(7.3\pm0.4)\times10^{-4}$ \massrate $\times\ (v_{exp}/50\,$\kms\!) using the canonical gas-to-dust ratio of 100.  Dividing our value for $r_{min} = 0.35$ pc by an assumed expansion velocity of 50 \kms gives an estimated expansion time of $\simeq$ 6800 yr and using our value of $r_{max} = 1.05$ pc, we estimate the mass-loss lasted for $\simeq 1.4\times10^4$ yr.  As noted in Table \ref{tab:MN_params}, the \twodust\ estimates for the dust mass and mass-loss rate are proportional to the distance squared and given the distance to MN 90 is unknown the values we have estimated in this work should be interpreted as hypothetical.

Based on the mid-IR morphology and results from radiative transfer calculations, MN 90 is a very luminous star surrounded by an optically thin dust shell located at about 4.8 kpc.  However, given the uncertainties in the distance and stellar parameters, no strong conclusions about the nature of MN 90 can be made.  It is necessary to observationally uncover the physical characteristics of the central star to determine the exact evolutionary status of MN 90.  Because MN 90 is not visible in the optical due to heavy extinction, near-IR spectroscopic information is needed to constrain the physical parameters for the central star.  Furthermore, continuous near-IR photometry to better characterize the variability of the source would be useful.  Although we interpret MN 90 as a candidate LBV star, it may be a B[e] supergiant or post-RSG star.  These classes have similar stellar and dust parameters as candidate LBVs and can be confused with each other.  Regardless of the exact evolutionary status, MN 90 seems highly likely to be an evolved, massive post-main sequence star.

\subsection{HD 168625}
The temperature map shows a higher temperature on the southern edge of the dust shell, $\sim$ 180 K, and a much lower temperature on the northern edge, $\sim$ 80 K.  This same temperature variation was found by \citet{OHara03}.  \citet{OHara03} suggest that one possible explanation for the different temperatures may be that the grain size distribution varies with respect to position in the nebula.  Because smaller grains tend to be warmer than larger grains, this would suggest that the smallest grains are in the southern shell with progressively larger grains towards the northern shell.  \citet{Pasquali02} interpret these distinctive optical morphologies as a variation in the gas-to-dust
mass ratio with respect to position in the nebula (i.e. it is higher in the south).

The smaller dust power-law emissivity ($\beta =  -0.33$) we have measured for HD 168625 compared to the value used by \citet[$\beta = 1.2$]{Robberto98b} is likely due to the fact that the circumstellar dust exhibits a large temperature variation (see Figure~\ref{HDTempMap}) and emission from small, transiently heated PAH grains (see \emph{IRAS} LRS in Figure~\ref{HD_sed}).  It is not realistic to model the dust as a single temperature--a temperature gradient would be more physically motivated--but in order to compare with previously measured dust temperatures of HD 168625 and to minimize the number of free parameters we only use a single modified blackbody.

\citet{Mahy16} used \emph{Herschel}/PACS spectroscopy and CNO abundances to estimate an initial mass of 28 -- 33 \solarmass for HD 168625 and propose that the star lost its mass during or just after the blue supergiant (BSG) phase and has not yet reached the red supergiant (RSG) phase.  Furthermore, they found that single star evolutionary tracks were able to explain the N content between the nebula and the central star.  This depends on the assumption that the star is a single star, whereas \cite{Smith15} argue that LBVs may be the product of binary evolution.  As mentioned in Section \ref{intro}, a wide-orbit binary companion to HD 168625 has been observed, but the influence this companion has on the evolution and morphology of HD 168625 is negligible.  Given that no X-ray emission has been observed \citep{Naze12} rules out mass transfer via Roche lobe overflow.  The rotation rate of HD 168625 is estimated to be 53 \kms \citep{Taylor14} which is not high enough to explain the bipolar structure based on rotation or binary merger unless the star passed through a super-Eddington phase.  Furthermore, no companion has been detected in radial velocity monitoring.
  
The photodissociation region (PDR) detected around HD 168625 \citep{Umana10, Mahy16} indicates that neutral gas makes up the majority of the shell's total mass.  \citet{Mahy16} measured a total ionized and neutral hydrogen gas mass of 1.17 \solarmass.  Using our dust mass estimate of $(2.5\pm0.1) \times 10^{-3}~M_{\sun}$, this corresponds to a gas-to-dust mass ratio of 470.  Using this gas-to-dust mass ratio, we estimate a total (gas plus dust) shell mass-loss rate of $(1.5\pm0.1)\times10^{-4}$ \massrate.  Dividing our value for $r_{min} = 0.06$ pc by the nebular expansion velocity measured by \citet{Pasquali02} gives an expansion time of $\simeq$ 2800 yr, and using our value of $r_{max} = 0.23$ pc, the mass-loss lasted for $\simeq 1\times10^4$ yr.

\section{Conclusion}
\label{conclusion}
Our SOFIA/FORCAST image at 37.1 \micron\ of MN 90 shows a limb-brightened, spherical dust shell surrounding the central star.  A least-squares fit of a $ B_{\lambda}(T_d)$ curve with emissivity $Q_{\lambda} \propto \lambda^{-0.76}$ to the SED of MN 90 yields a dust temperature of $59 \pm 10$ K, with the peak of the emission at 42.7 \micron.  Our \twodust\ model supports the idea that the dust resides in a thin, spherical dust shell and estimates that MN 90 lost $(3.2\pm0.1)\times10^{-2}$ \solarmass of dust in a massive stellar wind with a mass-loss rate of $(7.3\pm0.4)\times10^{-6}$ \massrate $\times\ (v_{exp}/50\,$\kms\!).  Using the canonical gas-to-dust mass ratio of 100, we estimate a total mass-loss of $3.2\pm0.1$ \solarmass\ for MN 90.  These \twodust\ estimates assume that MN 90 has a luminosity of $9.3 \times10^4~L_{\sun}$, at the lower end of luminosities of LBVs in their quiescent state, and a distance of 4.8 kpc.  Our \twodust\ model of MN 90 has good agreement with observations if we assume very small, transiently heated silicate grains.

Our SOFIA/FORCAST images between 7.7 -- 37.1 \micron\ of HD 168625 complement previously obtained mid-IR imaging.  The dust temperature map that we derive from our observations shows a temperature variation between the northern and southern shells, suggesting different grain size distributions between the two shells.  A least-squares fit of a $B_{\lambda}(T_d)$ curve with emissivity $Q_{\lambda} \propto \lambda^{0.33}$ to the SED of HD 168625 yields an estimated dust temperature of $170 \pm 40$ K, with the peak of the emission at 18.3 \micron.  Our detailed radiative transfer model using \twodust\  supports the claim that the dust resides in a thin, axisymmetric equatorial torus and estimates that HD 168625 lost $(2.5\pm0.1)\times10^{-3}$ \solarmass of dust in a massive stellar wind with a mass-loss rate of $(3.2\pm0.2)\times10^{-7}$ \massrate.  These \twodust\ estimates assume that HD 168625 has a luminosity of $4.5\times10^4~L_{\sun}$, at the lower end of luminosities of LBVs in their quiescent state, and a distance of 1.55 kpc. 

\bibliographystyle{apj}
\setcitestyle{notesep={: }}

\acknowledgments
The observations were made with the NASA/DLR Stratospheric Observatory for Infrared Astronomy (SOFIA). SOFIA is jointly operated by the Universities Space Research Association, Inc. (USRA), under NASA contract NNA17BF53C, and the Deutsches SOFIA Institut (DSI) under DLR contract 50 OK 0901 to the University of Stuttgart.  This research was supported by NASA under USRA funding for programs 02\_0101 and 03\_0131.  RDG was supported, in part, by the United States Air Force.

\vspace{5mm}
\facilities{SOFIA (FORCAST)}
\software{{\twodust}}

\bibliography{paperNotes}

\end{document}